\documentclass{aastex}

\usepackage{lscape}
\usepackage{epstopdf}
\usepackage{color}
\usepackage{epsfig}
\usepackage{comment}
\usepackage{amsmath}
\usepackage{graphicx}
\usepackage{lineno}

\slugcomment{accepted for the Astrophysical Journal}
 
\shorttitle{A 4~Kpc Molecular Gas Lane in Cygnus A}

\shortauthors{Carilli et al.}
 
\begin{document}

\title{A 4~Kpc Molecular Gas Lane in Cygnus A}

\author{
Christopher L. Carilli\altaffilmark{1},
Richard A. Perley\altaffilmark{1},
Daniel A. Perley\altaffilmark{2},
Vivek Dhawan\altaffilmark{1},
Roberto Decarli\altaffilmark{3},
Aaron S. Evans\altaffilmark{4,5},
Kristina Nyland\altaffilmark{6}
}

\altaffiltext{1}{National Radio Astronomy Observatory, P. O. Box 0,Socorro, NM 87801, USA, ccarilli@nrao.edu, ORCID: 0000-0001-6647-3861}

\altaffiltext{2}{Astrophysics Research Institute, Liverpool John Moores University, IC2, Liverpool Science Park, 146 Brownlow Hill, Liverpool L3 5RF, UK}

\altaffiltext{3}{INAF – Osservatorio di Astrofisica e Scienza dello Spazio di Bologna, via Gobetti 93/3, I-40129, Bologna, Italy}

\altaffiltext{4}{National Radio Astronomy Observatory, 520 Edgemont Rd, Charlottesville, VA}

\altaffiltext{5}{Astronomy Department, University of Virginia, 530 McCormick Road, Charlottesville, VA 22904 USA}

\altaffiltext{6}{U.S. Naval Research Laboratory, 4555 Overlook Ave SW, Washington, DC 20375, USA}

\begin{abstract}

We present the discovery of a 4 kpc molecular gas lane in the Cygnus A host galaxy, using ALMA CO 2-1 observations. The gas lane is oriented roughly perpendicular to the projected radio jet axis. The CO emission generally follows the clumpy dust lanes seen in HST I-band images. The total molecular gas mass is $30\times 10^8$ M$_\odot$ for Milky Way type clouds, and $3.6 \times 10^8$ M$_\odot$ for starburst conditions. There is a velocity change from the northern to southern CO peaks of about $\pm 175$~km~s$^{-1}$, and an apparently smooth velocity gradient between the peaks, although the emission in the central region is weak. In the inner $\sim 0.5''$ projected distance from the radio core, comparison of the CO velocities to those observed for H$_2$ 2.1218 $\mu$m emission shows higher velocities for the vibrationally excited warm molecular gas than the cooler CO 2-1 line emitting gas at similar projected radii. A possible explanation for these different projected velocities at a given radius is that the cooler CO gas is distributed in a clumpy ring at radius $\sim 1.5''$ to $2''$, while the warm H$_2$ 2.12$\mu$m emitting gas is interior to this ring. Of course, the current data cannot rule-out a clumpy, amorphous molecular gas distribution linearly distributed perpendicular to the radio jet axis. We consider surface brightness properties on scales down to $\sim 265$~pc, and discuss the Cygnus A results in the context of other radio galaxies with CO emission. 

\end{abstract}



\section{Introduction} \label{sec:intro}

Emission from molecular gas, via CO rotational lines, has been observed in a number of elliptical galaxies at the center of dense cooling flow clusters, often hosting extended radio jets \citep{edge01, SalomeCombes03}. The observed CO morphologies are varied, ranging from large scale filaments (tens of kpc-scales), possibly associated with gas cooling out of the hot phase \citep{salome06, salome11,olivares}, to gas associated with well defined kpc-scale dust lanes at the galaxy centers \citep{rose, hamer13, maccagni}, possibly representing a disk formed following a recent merger with a gas rich spiral. Molecular absorption toward the radio AGN is also seen on occasion, and optical depths can be high ($\tau \sim 1$). This absorbing gas possibly pertains to gas on small scales, $< 200$~pc, within the Bondi radius and possibly involved with fueling of the AGN accretion disk \citep{edge01, rose}.

Cygnus A, at $z = 0.056$, is the closest, by far, of the extreme luminosity radio galaxies ($L_{radio} > 10^{45}$ ergs s$^{-1}$), and hence has been extensively studied at all wavelengths \citep{cari96}. The giant elliptical host galaxy is at the center of a massive cooling flow, possibly merging, X-ray cluster \citep{halbesma, carilli94}. The nuclear regions within a few kpc radius of the core are a complex mixture of X-ray through radio continuum and line emitting structures, with the morphology at optical wavelengths being dictated by extensive dust lanes, as can be seen in the HST images \citep{jackson, tadhunter}. 

Cygnus A is considered a prime example of a Type II AGN \citep{barthel89}, in which our line of sight to the powerful AGN is highly obscured in the optical by a roughly edge-on dusty torus or disk in the inner tens to a hundred parsecs \citep{merlo, ueno, cari96, jackson, tadhunter}. In this scenario, the radio jet emerges along the axis of the obscuring torus. The latest VLBI analysis of Cygnus A suggest an inclination angle of the radio jet to the sky plane of $15.5^o$ \citep{boccardi16}, with the northwest jet approaching. The presence of an obscured broad line region is supported by the detection of scattered broad emission lines from the Cygnus A nuclear regions \citep{ogle97, antonucci94}.

\citet{lopez-rodriguez} present the nuclear spectrum integrated over a few arcsecond scales using observations in the far-IR with SOFIA and HERSCHEL, as well as low resolution submm data. Their spectral modeling yields a black body peak at $\sim 40\mu$m (see also \citet{radomski}), with an implied dust color temperature of 100~K to 150~K.  The total star formation rate for the Cygnus A host galaxy was estimated to be $\sim 10$ M$_\odot$ year$^{-1}$, based on low spatial resolution ($20''$) Spitzer, and other observations \citep{privon}, although this is highly uncertain due to the dominant contribution from thermal and non-thermal emission directly associated with the active nucleus. We return to a comparison with near-IR spectral lines in Section~\ref{sec:nearIR} below.

Millimeter studies of extended structures in the line and continuum in the inner few kpc of Cygnus A is complicated by the presence of the strong radio nucleus, close to 1 Jy, even in the millimeter, as well as much stronger emission from the radio lobes and hot spots situated in the skirts of the telescope primary beam. Broad HI 21cm absorption has been detected toward the nucleus of Cygnus A \citep{struve10}. Emission from near-IR H$_2$ vibrational emission lines has been observed, from warm molecular gas possibly heated by the AGN (\citet{ward}; the excitation temperature for H$_2$ vibration lines is $\ge 500$~K). To date, no detection has been made of emission from cooler molecular gas through e.g. rotational transitions of CO. \citet{evans05} single dish spectroscopy with a large ($\sim 1'$) beam, provides an upper limit to the CO 1-0 flux of $2.1\times 10^8$ K km s$^{-1}$ pc$^2$, although the result depends on assumed line width and residual spectral baseline subtraction accuracy. Likewise, no CO absorption has been detected towards the radio source to date \citep{barvainis94}. 

In this paper, we present the discovery of CO 2-1 emission over $\sim 4''$, across the nucleus of Cygnus A. We present the properties of this emission, and compare the results to other powerful radio galaxies in cooling flow clusters. We adopt the HI 21cm absorption redshift of 0.05634 as zero velocity \citep{struve10}, implying luminosity and angular diameter distances for Cygnus A of 253.4~Mpc, and 225~Mpc, respectively, using standard cosmology \citep{wright}. The latter implies that $1''~ =~ 1.1$~kpc.

\section{Observations and Reduction}\label{sec:obs}

We observed Cygnus A with ALMA for a total observing time of two hours, on July 9 and July 18, 2021, with a configuration with a maximum projected baseline of 3000~m, and projected minimum baseline of 15~m. However, the UV-coverage becomes sparse below about 30~m projected baselines, which corresponds to a fringe spacing of $\sim 10''$, or sensitivity to structures (source separations) of $\sim 5''$ \citep{monnier}. The short baseline coverage gives a rough indication of the largest scale structures to which we have good sensitivity, although the details will depend on the distribution of Fourier plane coverage relative to source structure in both size and orientation. Further, for spectral line imaging, the relevant maximum size scale is the size of the line emission per spectral channel, not the full source extent integrated over velocity. If there were diffuse line structures per channel that were larger than $\sim 5''$ in extent, our data would be decreasingly sensitive to such structures. For reference, the full source extent integrated over velocity is about $4''$, while the largest structures in the channel maps at 10 km s$^{-1}$ channel$^{-1}$ are about $1.5''$ in maximum extent. 

Two of the 1.87 GHz sub-bands were overlapped to cover the CO 2-1 line. The summed spectrum is centered at an LSRK redshift of z = 0.05637 (LSRK frequency of 218.2415 GHz). We adopt this redshift as zero velocity, corresponding to within 5 km s$^{-1}$ of the HI 21cm absorption line velocity \citep{struve10}. The other two sub-bands were centered at 232.8 GHz and 234.7 GHz for continuum observations. 
A measurement set with NRAO pipeline spectral line calibration performed was delivered by the  North American ALMA Science Center.\footnote{\url{https://help.almascience.org/kb/articles/where-can-i-get-additional-information-for-my-na-added-value-data-products}}
The flux scale and bandpass calibrator was J2253+1608, and the phase calibrator was J2007+4029. The radio core of Cygnus A had a measured flux density of 0.53 Jy at 233 GHz, and is unresolved. This flux density is high enough to allow for self-calibration. The spectral data were averaged to 3.9 MHz channel$^{-1}$, and in time to 4s, and exported to AIPS for self-calibration. For each day, and sub-band, one iteration of phase-only self-calibration was performed, averaging solutions across the band and using a solution averaging time of 3min. The spectral line data were then exported back to CASA for imaging. 

In CASA, UVCONTSUB was used to subtract the continuum from the visibilities by fitting a linear baseline in frequency, and adopting frequency ranges that avoided the channels with line signal. Spectral image cubes were synthesized using TCLEAN, using a channel width of 7.3 MHz (10.0 km s$^{-1}$).  Briggs weighting with Robust = 2, was used, and the images were cleaned down to a residual threshold of 0.45 mJy beam$^{-1}$ channel$^{-1}$. A residual continuum subtraction was performed using the spectral image cubes with IMCONTSUB, again assuming a linear baseline. This latter step was done over a wide spectral range to remove residual continuum at the position of the bright radio nucleus due to residual bandpass errors (see Section~\ref{sec:results}). The synthesized beam has a FWHM = $0.31''\times 0.17''$, major axis position angle = $-24^o$, and the rms noise per channel is $\sim 0.33$ mJy beam$^{-1}$.


\section{Astrometry}\label{sec:astrometry}

\begin{figure*}
\centering
\includegraphics[trim=0.5in 5in 1in 0.5in, clip, width=1.1\linewidth]{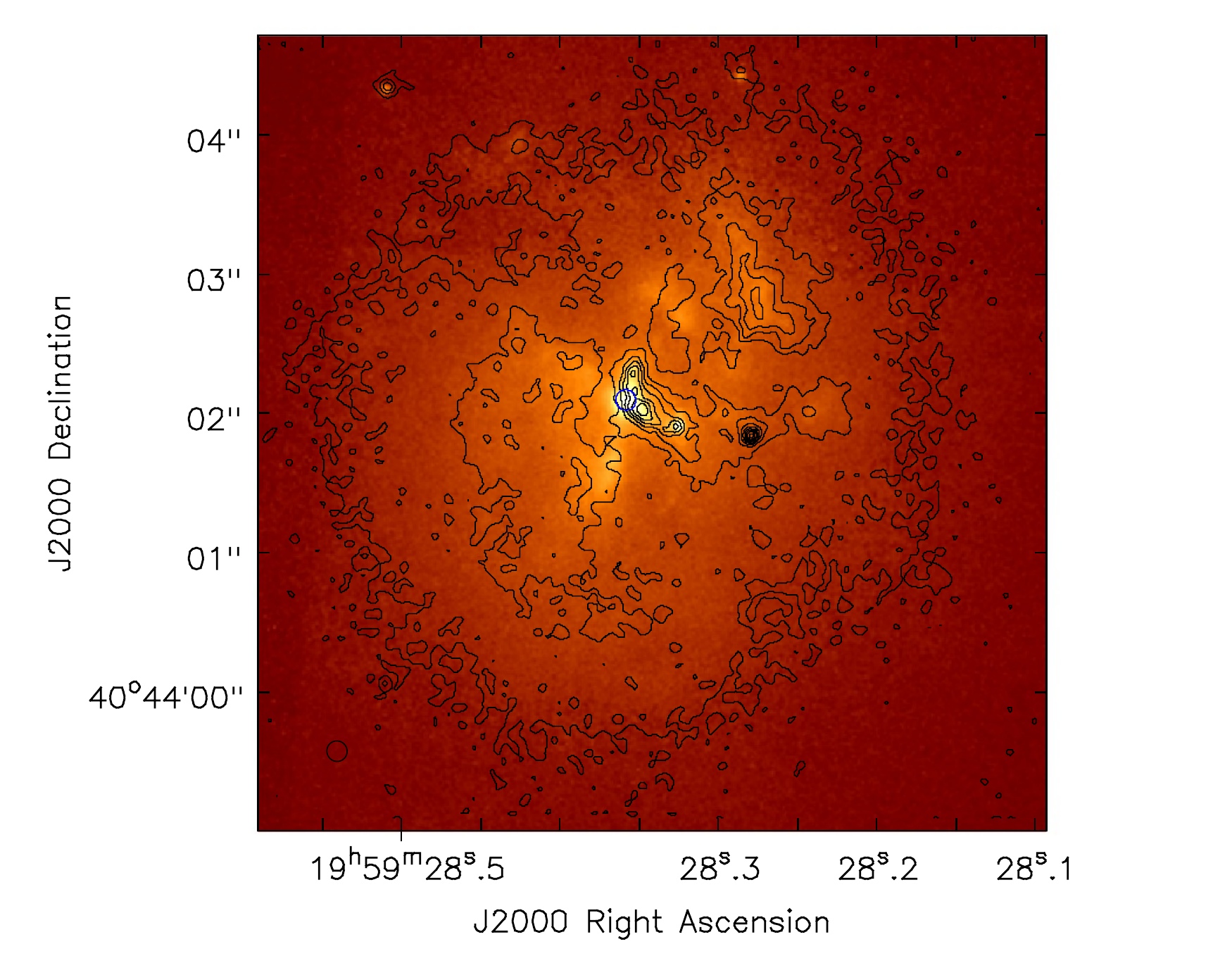}
\caption{Keck $K'$ band image of the Cygnus A galaxy in color \citep{canalizo}, with a resolution FWHM $\sim 0.085''$, plus HST I band image in contours \citep{jackson}, with a resolution FWHM $\sim 0.18''$. The blue circle shows the position of the radio nucleus at 11 GHz (which has a resolution of $0.15''$; core position (J2000) = 19h59m28.358s+40$^o$44$'$02.01$''$). The two astrometric stars are located on the upper edge of the frame, while the third astrometric source (the radio transient and associated persistent optical source), is $0.4''$ southwest of the nucleus. }
\label{fig:KeckHST}
\end{figure*}

We make comparisons of the CO emission to HST I band and Keck $K'$ band images, hence relative astrometry is important. We employ the HST I-band image from \citet{jackson}, and the Keck adaptive optics $K'$-band image from \citet{canalizo} (Figure~\ref{fig:KeckHST}). Note that the VLA and ALMA observations employed the same phase calibrator (J2007+4029), which sets the astrometry between the millimeter to centimeter images to very high accuracy (few mas or better). This relative mm-cm astrometry was confirmed through the position of the strong radio nucleus.

There are two reasons why comparing features observed in the Cygnus A galaxy from multi-wavelength observations is difficult. First, Cygnus A is at low Galactic latitude ($\sim 5^o$), seen through the dense, confused Cygnus region of the Galaxy. Second, the optical AGN of Cygnus A is highly obscured, with $A_v > 50$, based on near-IR spectroscopy \citep{merlo}, HI 21cm absorption \citep{struve10}, and X-ray absorption \citet{ueno}. Previous studies using HST optical images simply aligned the optical peak with the radio core (e.g. \citet{jackson}). As we shall see, this alignment is incorrect.

In their near-IR image, \citet{canalizo} find a second relatively bright, compact source about $0.4''$ southwest of the radio core. The interpretation of this second near-IR peak remains uncertain: perhaps a young dense star cluster, or a secondary AGN. Recent discovery of a radio transient associated with this secondary peak by \citet{perley17}, supports the secondary AGN hypothesis for Cygnus A, although not conclusively (see also \citet{devries}). For this paper, the important result is that this second radio-optically aligned source allows for improved radio-optical astrometry within $2''$ of the Cygnus A nucleus, as follows.  

First, we align the radio and the Keck near-IR images using the primary nucleus (which can be seen, although attenuated, at $K'$), and the transient, as was done in \citet{perley17}, for which they find relative agreement for the dual alignment of 10~mas for the two sources. In all cases, we adopt the plate scale and rotation of the original images, and only adjust the RA and Dec, and we adjust all images back to the radio defined astrometry.

We then align the HST I-band and Keck near-IR images. The difficulty here is that we cannot use the primary nucleus, due to obscuration in the optical. For this, we use two stars about $2''$ northeast and northwest of the core, plus the secondary nucleus that is seen in both the optical and near-IR. Adjusting astrometry for all three sources, we find agreement to within $\sim 20$~mas between sources. 

Note that the astrometric agreements quoted above are not statistical, but simply represent the mean differences observed between the positions of the few sources used for alignment for images in different bands. The spatial resolution of the Keck adaptive optics image is quoted as $\sim 0.05''$ \citep{canalizo}, while we derive source sizes from Gaussian fitting to a few point sources in the HST I band image of $\sim 0.18''$. A relative astrometric accuracy of $\sim 0.2''$ is easily enough for a meaningful morphological comparison of the CO and HST optical emission below. 

For reference, and for potential future high resolution imaging studies, Figure~\ref{fig:KeckHST} shows the astrometrically aligned HST and Keck images, with the location of the radio nucleus at 11 GHz also indicated. 

\section{Results}\label{sec:results}

\begin{figure*}
\centering
\includegraphics[trim=1.8in 1.7in 0.3in 1in, clip, width=1.4\linewidth]{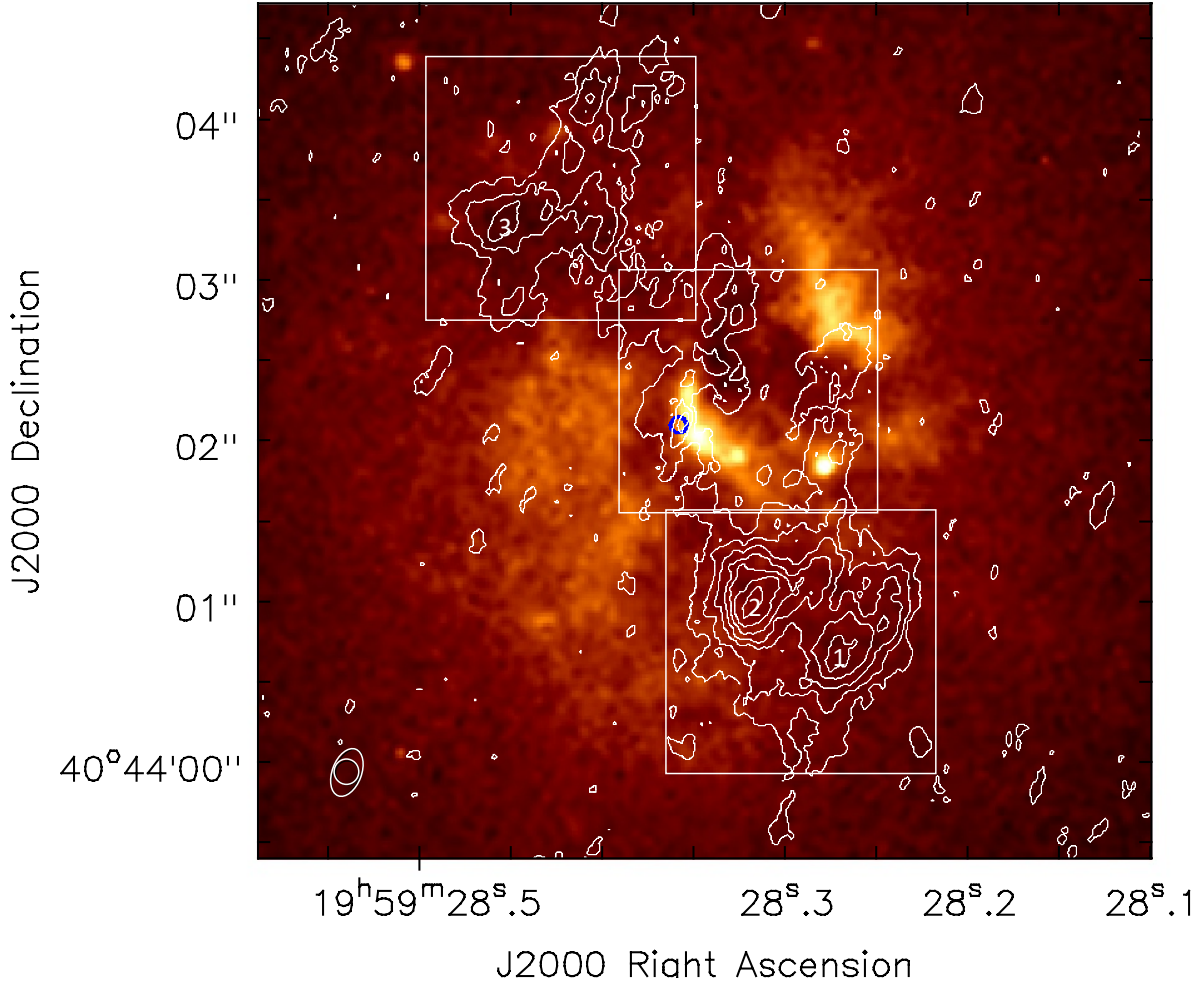}
\caption{HST I band image with the ALMA CO 2-1 moment 0 map (velocity integrated emission), as contours with a resolution of $0.31''\times 0.17''$, major axis position angle = $-24^o$. Contour levels are: 0.033, 0.066, 0.099, 0.132, 0.165, 0.198, 0.231, 0.264 Jy beam$^{-1}$ km s$^{-1}$. The blue circle shows the position of the radio nucleus at 11 GHz. The boxes show the regions over which the CO emission was integrated for Figure~\ref{fig:SpecInt}. The numbers (1,2,3) indicate the positions where spectra were extracted for individual beams in Figure~\ref{fig:Specpoint}. The width of the cut used for the position-velocity plot in Figure~\ref{fig:PV} is $1.3''$, with a position angle of $+32^o$ (measured counterclockwise from North), and centered on the radio core. This cut encompasses the emission within the three boxes shown.
}
\label{fig:HSTCO}
\end{figure*}

CO 2-1 emission has been detected with these ALMA observations in the velocity range $-400$ km s$^{-1}$ to $+250$ km s$^{-1}$.  Figure~\ref{fig:HSTCO} shows the velocity integrated emission (moment 0) image over this velocity range, generated using the CASA task IMMOMENTS, along with the HST image. A clipping level of $0.6$ mJy beam$^{-1}$, was employed, meaning only signal above this value was included in the moment sums.  CO emission is seen extending over about $4''$ ($\sim 4.4$~kpc), oriented roughly north-south, and generally following the dust lanes seen on the HST images. The two dominant emission regions are about $1.5''$ in size, located $\sim 1.5''$ to $2''$ northeast and southwest of the nucleus. Faint emission extends across the center of the galaxy. Generally, the emission is clumpy, with a significant asymmetry from the bright, double-peak emission to the south, and the more diffuse emission to the north. 

\begin{figure*}
\centering
\includegraphics[trim=0.8in 2.2in 0in 0in, clip, width=1.05\linewidth]{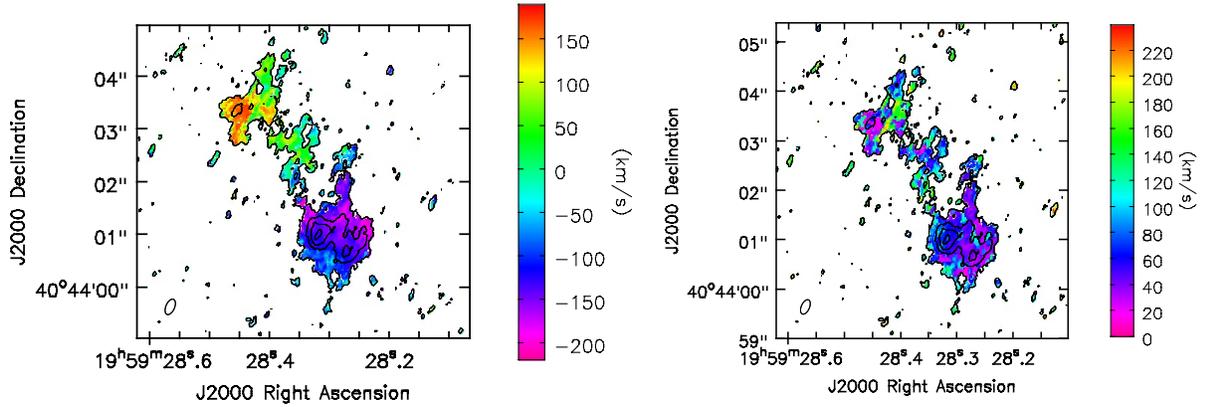}
\caption{{\bf Left:} Color scale is the moment 1 image (intensity weighted mean velocity) for the CO 2-1 emission. {\bf Right:} Color scale is the CO 2-1 velocity dispersion. The contours in both cases are the moment 0 map (Figure~\ref{fig:HSTCO}). The resolution is $0.31''\times 0.17''$, major axis position angle = $-24^o$.
}
\label{fig:Mom}
\end{figure*}

Figure~\ref{fig:Mom} shows the mean velocity and velocity dispersion for the CO emission, again using a clipping level of $0.6$ mJy beam$^{-1}$. The emission shows a gradient north-south between the two dominant emission regions, from about $+150$ km s$^{-1}$ in the north, to about $-200$ km s$^{-1}$ in the south. The velocity dispersion in the high signal to noise regions varies from 16 km s$^{-1}$ to 60 km s$^{-1}$. Some higher velocity dispersions are seen, $\ge 100$~km s$^{-1}$, but these are in lower signal-to-noise regions, where noise peaks above the clipping level in a single off-line channel can significantly perturb the dispersion measurement. 

\begin{figure*}
\centering
\includegraphics[trim=0.6in 3in 0in 3in, clip, width=1.\linewidth]{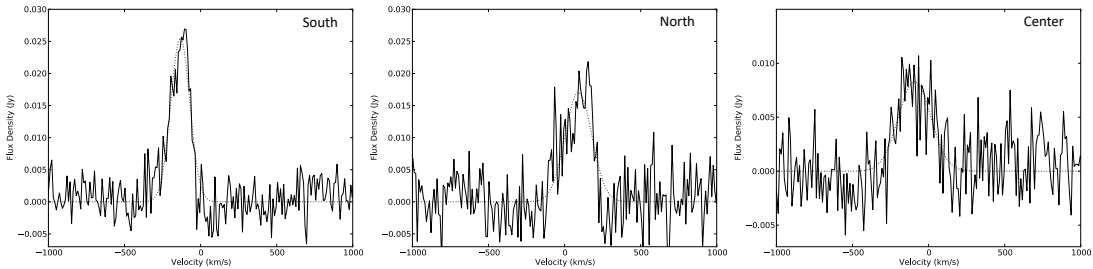}
\caption{Spectra of the integrated CO 2-1 emission from three regions of $\sim 1.5''$ in size (Figure~\ref{fig:HSTCO}), and channel width of 10 km s$^{-1}$. Dotted lines are the Gaussian fits to the data. 
}
\label{fig:SpecInt}
\end{figure*}

Figure~\ref{fig:SpecInt} shows spectra summing over regions $\sim 1.5''$ in size, centered on the northern, central, and southern CO emitting regions, along with Gaussian fits to the spectra. The regions employed are delineated by boxes in Figure~\ref{fig:HSTCO}. Values for the resulting fit parameters for line peak, mean velocity, FWHM, and velocity integrated line flux are listed in Table 1. Note that, for a Gaussian profile, line FWHM $= 2.355\times$ velocity dispersion.  Figure~\ref{fig:Specpoint} shows point-spectra at full spatial resolution (i.e. in Jy beam$^{-1}$) for the three peaks in the moment zero image, as indicated in Figure~\ref{fig:HSTCO}. Values for the fit parameters are given in Table 2. 

The spectrum of the radio core itself (again, $\sim 0.5$~Jy continuum strength), shows a broad ripple, with positive and negative structure on frequency scales $\ge 1$~GHz, or $\ge 1000$ km s$^{-1}$, at $+/-0.5$ mJy beam$^{-1}$, implying a spectral dynamic range of $\sim 1000$, over the full bandpass. No absorption is seen toward the core at this level, at a resolution of 10 km s$^{-1}$ channel$^{-1}$. 


\section{Analysis}\label{sec:analysis}

\subsection{Molecular Gas Masses}

The observed CO 2-1 emission generally follows the north-south oriented filaments of dust obscuration seen on the HST image in the inner $4''$ of the Cygnus A galaxy, although the correspondence is not exact. We consider the physical properties of this emission based on the CO spectra. 

Table 1 lists the velocity integrated CO 2-1 fluxes in Jy km s$^{-1}$ derived from Gaussian fitting to the line profiles. Table 2 lists the results from Gaussian fitting to the point-position spectra (Figure~\ref{fig:Specpoint}), in Jy beam$^{-1}$ km s$^{-1}$. 

We then calculate $L'_{2-1}$ in K km s$^{-1}$ pc$^2$, corresponding to the areal integrated source brightness temperature, using equations in Section 2.4 in  \citet{carilliwalter}: $L'_{2-1} = 3.25\times 10^7~S_{2-1}\Delta v ~D_L^2~ (1+z)^{-1}~ \nu_{r,2-1}^{-2}$, where $S_{2-1}\Delta v$ is the CO 2-1 velocity integrated flux in Table 1 in Jy km s$^{-1}$, $D_L = 253.4$~Mpc, $z = 0.05637$, and $\nu_{r, 2-1} = 230.538$~GHz = the rest frequency of the CO 2-1 line. $L'$ relates to the integrated line luminosity, $L$, as:  $L = 3\times 10^{-11}~ \nu_{rest}^3~ L'$~L$_\odot$.

To derive molecular gas masses using standard conversions, the $L'_{2-1}$ values need to be extrapolated to CO 1-0 fluxes, for which an excitation is required. We adopt two different values for the $L'_{2-1}/L'_{1-0}$ ratio: a value of 0.5 appropriate for typical star forming galaxies, such as the Milky Way (MW), and a value of 0.85 appropriate for denser, warmer gas in extreme starburst galaxies (SB) (see Table 2 in \citet{carilliwalter}). Previous surveys of CO emission from elliptical galaxies find sub-thermal excitation, comparable to that seen for Milky Way clouds \citep{vila-vilaro03}, or, in one case, an excitation in-between Milky Way and starburst \citep{vila-vilaro19}, although these were typically lower mass ellipticals with much lower luminosity AGN. Conversely, \citet{rose} find a high excitation in the disk in Hydra A consistent with constant brightness temperature, i.e. closer to SB values. 

The conversion from $L'_{1-0}$ to molecular gas mass also requires an empirically derived conversion factor, $\alpha_{CO}$ in  $\rm M_\odot/(K~km~s^{-1}~pc^2$). This conversion factor is different for starburst vs. normal star forming galaxies \citep{bollato, carilliwalter}. For a Milky Way type galaxy, we adopt $\alpha_{CO} = 4.0$, and for a starburst, $\alpha_{CO} = 0.8$ \citep{carilliwalter, bollato}. 

We note that a recent study of the molecular disk in Centaurus A \citep{miura}, an elliptical galaxy with a prominent dust and gas lane, found a MW conversion factor for clouds beyond 200~pc from the nucleus, but possibly a factor $\sim 2.5$ higher conversion factor for the circumnuclear disk (CND) on smaller scales. \citet{miura} suggest this higher conversion factor (meaning, more H$_2$ mass for a given CO luminosity), could be due to a higher radiation or cosmic ray field near the AGN. For Cygnus A, the emission we observe is mostly at much larger radii ($> 1$~kpc), than the CND in Centaurus A. On the other hand, the obscured Cygnus A AGN may be 100 times, or more, higher luminosity than that in Centaurus A, based on hard X-ray observations \citep{israel,cari96}. If the conversion factor was higher than MW, the molecular gas masses would increase accordingly. 

Table 1 lists the molecular gas masses under the MW and SB assumptions, for the three regions defined in Section~\ref{sec:results}. Summing over the three regions gives $L'_{2-1} = 3.8\times 10^8$ K km s$^{-1}$ pc$^2$. \citet{evans05} derive an upper limit of $L'_{1-0} < 2.1\times 10^8$ K km s$^{-1}$ pc$^2$, although, again, this limit was based on very low spatial resolution single dish spectroscopy with admittedly significant spectral baseline structure.  Taken at face-value, the \citet{evans05} limit would imply a modestly super-thermal excitation for CO(2-1). This result needs to be checked with better CO(1-0) observations. 

Summing over the three regions, the total molecular gas mass  is $30.3\times 10^8$~M$_\odot$ for the MW case, and $3.6\times 10^8$~M$_\odot$ for the SB case. 

For the surface brightness spectra at a given position at full spatial resolution ($0.31''\times 0.17''$), the velocity integrated surface brightness in Jy beam$^{-1}$ km s$^{-1}$, can be converted into velocity integrated line brightness temperature $W(CO)$ in K km s$^{-1}$, using the beam size and wavelength \citep{bollato}. These values are listed in Table 2 for the three brightest CO emission positions in Figure~\ref{fig:HSTCO}.  Also listed in Table 2 are the implied $L'$ values and molecular gas masses within each beam area, again assuming both MW and SB conversions.

\subsection{Velocities}

Figure~\ref{fig:PV} shows the position-velocity plot for the CO emission for a slice along the major axis of the emission (PA $= +32^o$; PA measured counterclockwise from North), with a width of $1.3''$. The emission is dominated by the northern and southern peaks, with faint emission extending across the central regions at the $\sim 2$ to $3\sigma$ level. There is a velocity gradient across the source from North to South. Such a gradient could be due to rotation, infall, or outflow, or just stochastic chance. The current data are insufficient to differentiate between the various possibilities.

Under the assumption that the velocity gradient from north to south is due to rotation, and assuming the rotating gas is close to edge-on, as considered the hypothesized inner obscuring torus or disk on smaller scales (See Section~\ref{sec:intro}), then the implied gravitational mass inside 2~kpc radius would be $\sim 2\times 10^{10}$~M$_\odot$. For reference, the black hole mass in Cygnus A has been estimated from near-IR spectroscopy to be $\sim 2.5\times 10^9$ \citep{tadhunter03}. 

\subsection{Comparison to near-IR lines}\label{sec:nearIR}

\citet{riffle21} presents IFU imaging of multiple spectral lines in the near-IR from the inner $\sim 1.5''$ of the Cygnus A galaxy. The near-IR lines include Pa$\alpha$, vibrational lines of warm, dense H$_2$, and high ionization `coronal lines' of Si and others, with a spatial resolution of $\sim 0.2''$. They confirm at high resolution the bi-conical structure, with the cone axis along the radio jet axis, and a half opening angle of $45^o$ \citep{tadhunter03,jackson,canalizo}, most clearly seen in the Pa$\alpha$ image. They find evidence for hot, coronal line emission within the bi-cone area along the radio jet, to a distance of at least 1~kpc, with possible outflow velocities up to 600 km s$^{-1}$ within 0.5~kpc radius. 

All the near-IR lines have high velocity dispersions, between $50$ km s$^{-1}$ to 350 km $^{-1}$. These higher dispersions may indicate stronger local feedback with decreasing radius from the active core region, either hydrodynamic or radiative. Analysis of H$_2$ line ratios implies a gas temperature in the inner $\sim 0.5''$ of $> 1000$~K, likely due to X-ray heating by the AGN (see also \citet{wilmer}).

Most relevant to the CO results presented herein is the observation of a similar north-south velocity gradient perpendicular to the projected radio jet direction, seen most clearly in the H$_2$ 2.12~$\mu$m image. For the H$_2$ 2.12~$\mu$m line, the surface brightness peaks at a radius $\sim 0.31''$ $= 0.35$~kpc. The projected velocity also peaks at this position, with a value of $\pm150$~km~s$^{-1}$ (see Figure 6 in \citet{riffle21}). Comparing these values to the  CO velocity field, the CO has a much lower velocity at this radius, by at least a factor two  (Figure~\ref{fig:PV}). The CO only reaches a maximum velocity of $\pm 175$ km s$^{-1}$, at $2''$ distance from the core. In other words, the H$_2$ 2.12~$\mu$m velocities imply a very rapidly rising velocity field with distance from the core, while the observed CO velocity gradient is more gradual. The implication is that the CO 2-1 and H$_2$ 2.12~$\mu$m emission projected in the inner $\sim 0.5''$ of Cygnus A may not be physically co-located.

One possible explanation is that the cooler molecular gas traced by the CO 2-1 emission is not in a continuous disk across the center of the galaxy, but weighted toward a ring structure at $\sim 1.5''$ to $2''$ radius, while the warmer molecular gas traced by H$_2$ 2.12~$\mu$m emission is interior to this ring. Such a ring would be consistent with the projected column density peaks being at the extremities for the CO (Figure~\ref{fig:HSTCO}). The smooth CO velocity gradient from north to south would then be due to projection. 

We emphasize that the CO emission across the center of the source is weak, and generally clumpy in appearance. Any firm conclusions on spatial and velocity distribution of the molecular gas across the inner regions requires more sensitive observations.

\subsection{Spectra at Full Spatial Resolution}

\begin{figure*}
\centering
\includegraphics[trim=0.6in 3in 0in 3in, clip, width=0.95\linewidth]{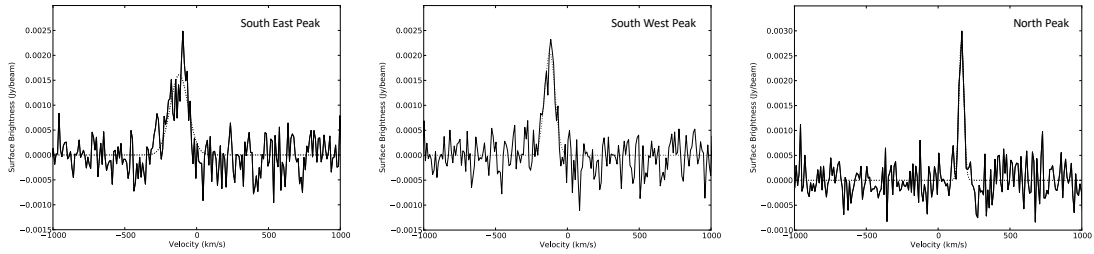}
\caption{Spectra of the CO 2-1 surface brightness from three peaks in the northern and southern components (southwest = 1, southeast = 2, north = 3 in Figure~\ref{fig:HSTCO} and Table~\ref{tab:peaks}). The resolution is $0.31''\times 0.17''$, major axis position angle = $-24^o$, and channel width of 10 km s$^{-1}$. Dotted lines are the Gaussian fits to the data. 
}
\label{fig:Specpoint}
\end{figure*}

\begin{figure*}
\centering
\includegraphics[trim=0.0in 5.2in 0in 0in, clip, width=1.05\linewidth]{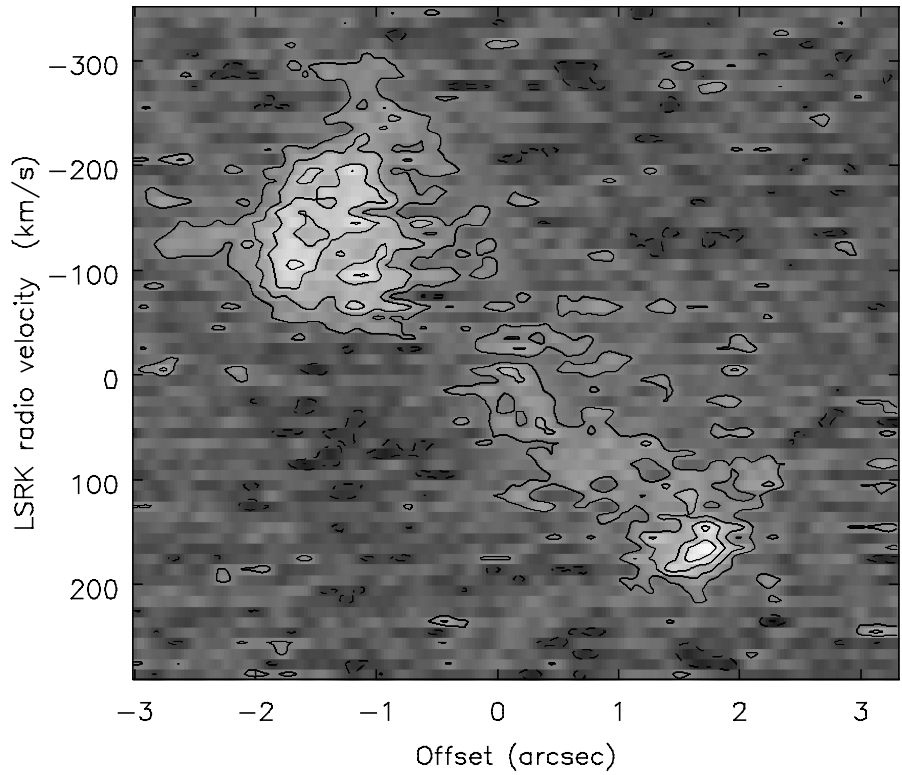}
\caption{A position-velocity plot for the CO 2-1 emission for a cut along the major axis (PA $= +32^o$), with a width of $1.3''$. 
Zero velocity is defined at LSRK redshift of z = 0.05637. The brightness values are averages across the width, with a PV image rms of $0.16$ mJy beam$^{-1}$. The contour levels are: -0.48, -0.24, 0.24, 0.48, 0.72, 0.96 mJy beam$^{-1}$.
}
\label{fig:PV}
\end{figure*}

At full spatial resolution ($0.31''\times0.17''$ = 340~pc $\times$ 190~pc), Gaussian fitting to spectra at the three CO peak emission positions in Figure~\ref{fig:HSTCO}, show a minimum velocity dispersion of 17 km s$^{-1}$ and molecular gas mass of $3.6\times 10^7$ M$_\odot$ for MW conditions or $0.42\times 10^7$ M$_\odot$ for SB conditions (position 3 North in Table 2). The maximum dispersion is 62 km s$^{-1}$ and molecular gas mass of $7.5\times 10^7$ M$_\odot$ for MW conditions or $0.88\times 10^7$ M$_\odot$ for SB conditions (position 2 southeast in Table 2, although the line profile is not particularly Gaussian, see Figure~\ref{fig:Specpoint}). 

Of particular interest is the study of cloud properties on GMC-scales ($\le 100$~pc), such has been done in nearby galaxy samples. However, a proper comparison requires significantly higher spatial resolution: the recent study of the molecular clouds in the disk in Centaurus A by \citet{miura} was done at 20~pc resolution (factor 140 smaller beam physical area than for Cygnus A), while the PHANGS molecular survey of nearby galaxies employed 100~pc resolution \citep{leroy} (factor seven smaller beam area). At the least, we can extrapolate the relations observed for nearby galaxies to individual peaks in Cygnus A. Also keep in mind that, given the shortest baselines of our observations, we are insensitive to diffuse line emission larger than $\sim 5''$ per 10 km s$^{-1}$ channel$^{-1}$ (Section~\ref{sec:obs}).

\citet{miura} present the Line Width - Size relation \citep{larson}, for the Centaurus A clouds in comparison with other samples, for which the maximum measured cloud radii $\sim 100$~pc. \citet{miura} employ a MW gas mass conversion factor. Extrapolating from their relationship seen for the Centaurus A clouds to our resolution $\sim 265$~pc (= mean of major and minor axes), leads to an expected molecular gas mass of $\sim 4\times 10^7$ M$_\odot$ for a MW conversion, with a factor $\sim 3$ scatter for the Centaurus A clouds. This mass is comparable to the cloud masses in Table 2 for Cygnus A assuming MW conditions, but an order of magnitude larger than the SB mass values. We emphasize that our physical resolution is not adequate to isolate individual GMCs, but more likely GMC complexes, for which the extrapolation to larger scale above is suspect.

We also consider the relationship between cloud velocity dispersion and surface density, derived from CO 2-1 emission in the PHANGS sample by \citet{sun} (although, again, PHANGS being at higher spatial and spectral resolution). The PHANGS survey also employs the MW conversion factor. From the point spectra at the peak CO surface brightness positions in Cygnus A (Figure~\ref{fig:Specpoint}), the mean velocity dispersion is $\sim 40$ km s$^{-1}$, and the mean molecular gas mass surface density is $\Sigma_{mol} \sim 765$ M$_\odot$ pc$^{-2}$ for MW conditions. Comparing the mean Cygnus A values to the ensemble distribution of cloud properties in Figure 1 of \citet{sun}, the Cygnus A values for velocity dispersion vs. surface mass density are consistent with the PHANGS distribution, with both quantities being at the the high end of the distribution, at least when using MW conversions. For SB conversion, the surface mass density would be an order of magnitude lower, and fall outside the PHANGS distribution.

\section{Discussion}\label{sec:discussion}

A recent compendium of CO emission from radio galaxies using archival ALMA data, as well as data from the literature, shows that, for high luminosity radio galaxies, about 40\% are detected in CO emission \citep{audibert}. This inhomogeneous sample spans a redshift range of $z \sim 0.1$ to 2.5, and molecular gas masses from $2.5\times 10^{9}$ M$_\odot$ to $3\times 10^{11}$ M$_\odot$. Likewise, a sample of giant ellipticals in cooling flow clusters at $z < 0.5$, had a 60\% detection rate in CO emission, with molecular gas masses in the range of $5\times 10^8$ to $3\times 10^{11}$ M$_\odot$ \citep{edge01}. In both cases, Cygnus A is at the low end of the molecular gas mass distribution for the detected systems. 

A few nearby BCG elliptical galaxies, often with radio jets, have been imaged in CO emission on kpc, or larger, scales.  Many show morphologies consistent with molecular gas cooling out of the warmer ISM phase on scales of tens of kpc or greater, such as NGC 1275 (Perseus A), Abell S1101, RXJ1939.5,  and NGC 4696. Evidence includes alignment of the molecular gas with the edges of the X-ray cavities seen in the cluster, possibly indicating gas compression by the expanding radio source \citep{salome06, salome11,olivares}. The gas in these cases also often aligns with warm ionized gas seen in H$\alpha$ emission \citep{olivares}.  

A few such galaxies show regularly rotating molecular gas disks, perpendicular to the radio axis of the jet, such as NGC 262, NGC 5128 (Centaurus A), and Hydra A \citep{olivares}. In these cases, the disk origin may be the result of recent accretion of a gas rich spiral \citep{rose, hamer13, maccagni, olivares}. 

The CO 2-1 emission in Cygnus A is extended roughly north-south from the radio core, perpendicular to the radio jet axis, and generally aligned with the dust lanes to about $2''$ radius. The emission shows a clear velocity gradient, possibly suggesting a disk. But the emission is very clumpy, heavily weighted toward extreme radii, and asymmetric. Comparison with the velocity field of the H$_2$ 2.12~$\mu$m line emission at $\le 0.5$~kpc projected radius suggests the CO 2-1 emission may come from cooler molecular gas at larger physical radii than the warm, vibrationally excited $H_2$ 2.12~$\mu$m line emission, with the CO emission perhaps coming from gas in a clumpy ring (Section~\ref{sec:nearIR}). A similar relative morphology for CO 3-2 and $H_2$ 2.12~$\mu$m emission, i.e. the inner hole in CO, filled by the $H_2$ emission, has been seen in Centaurus A, although on somewhat smaller scales (400~pc diameter for the CO ring, and 40~pc diameter for the central $H_2$ emission; \citet{espada}).

In Cygnus A, there is also evidence from low spatial resolution ($1.3''$) observations of the H$\alpha$ emission that the warm ionized gas is extended north-south from the nucleus, to a radius $\sim 10$ kpc. This H$\alpha$ emission projects along the inner edge of the radio lobe seen at 1.4 GHz \citep{carilli89}. \citet{carilli89} suggest that this north-south alignment of the H$\alpha$ emission and the inner edge of the radio lobe, could be gas compression by the back-flowing radio lobe, or simply a chance projection. 

Overall, the ALMA observations have revealed, for the first time, CO rotational line emission from Cygnus A. However, the data are inadequate to conclude whether the molecular gas is in a disk or ring, or just  clumpy, amorphous, but linearly extended gas lanes. We note that the Cygnus A cluster cooling flow rate within 90 kpc of the cluster center has been estimated to be 250 M$_\odot$ year$^{-1}$ \citep{reynolds96,edge01}. This rate would be adequate to build the observed molecular gas in a few $\times 10^6$ years.

These observations accentuate the complexity of the inner regions of the closest powerful radio galaxy, representing one more constituent in the rich ISM, which includes a complex distribution of stars, dust, ionized and neutral atomic gas, and warm, and now cool, molecular gas. The region presents an ideal opportunity to study the interplay of the various ISM constituents on sub-kpc scales in an extreme luminosity radio galaxy \citep{nyland}.  

\acknowledgments
The National Radio Astronomy Observatory is a facility of the National Science Foundation operated under cooperative agreement by Associated Universities, Inc. We thank Fabian Walter for useful discussions. This paper makes use of the following ALMA data from program 2019.1.00894.S. ALMA is a partnership of ESO (representing its member states), NSF (USA) and NINS (Japan), together with NRC (Canada), MOST and ASIAA (Taiwan), and KASI (Republic of Korea), in cooperation with the Republic of Chile. The Joint ALMA Observatory is operated by ESO, AUI/NRAO and NAOJ. Basic research in radio astronomy at the U.S. Naval Research Laboratory is supported by 6.1 Base Funding.

\begin{landscape}
\begin{table}
\centering
\scriptsize
\caption{Gaussian Fitting Results to Spectra of Total Emission of Three Areas \label{tab:regions}}
\begin{tabular}{lccccccc}
\hline\hline
Region & Gaussian Peak & Velocity & FWHM & Integrated Flux & $L'_{CO2-1}$ & M$_{gas,MW}$ & M$_{gas,SB}$ \\ 
~ & mJy & km s$^{-1}$ & km s$^{-1}$ & Jy km s$^{-1}$ &   K km s$^{-1}$ pc$^2$ & M$_\odot$ &  M$_\odot$ \\ 
~ & ~ & ~ & ~ & ~ & $\times 10^7$  & $\times 10^8$  & $\times 10^8$ \\
\hline
South & $25\pm 1.1$ & $-132\pm 3$ & $147\pm 8$ & $4.0\pm 0.3$ &   $14.9\pm 1.0$ & $12.0\pm0.8$ & $1.4\pm0.1$ \\
Center & $8.3\pm 0.8$ & $-92\pm 11$ & $250\pm 26$ & $2.2\pm 0.3$ &   $8.2\pm 1.2$ & $6.6\pm 0.9$ & $0.8\pm0.1$ \\
North & $17\pm 1.3$ & $90\pm 8$ & $212\pm 19$ & $3.9\pm 0.4$ &     $14.6\pm 1.7$ & $11.7\pm 1.3$ & $1.4\pm 0.2$ \\
\hline
\end{tabular}
\end{table}

\begin{table}
\centering
\scriptsize
\caption{Gaussian Fitting Results to Spectra of Peak Surface Brightnesses \label{tab:peaks}}
\begin{tabular}{lcccccccc}
\hline\hline
Position & Gaussian Peak  & Velocity & FWHM & Integrated Flux & W$_{CO 2-1}$ & $L'_{CO 2-1}$ & M$_{gas,MW}$ & M$_{gas,SB}$ \\ 
~ & mJy beam$^{-1}$ & km s$^{-1}$ & km s$^{-1}$ & Jy beam$^{-1}$ km s$^{-1}$ & K km s$^{-1}$ & K km s$^{-1}$ pc$^2$ & M$_\odot$ & M$_\odot$ \\ 
~ & ~ & ~ & ~ & ~ & ~ & $\times 10^6$ & $\times 10^7$ & $\times 10^7$ \\ 
\hline
1 Southwest & $2.1\pm 0.2$ & $-122\pm 3$ & $86\pm 8$ & $0.19\pm 0.02$ & $99\pm 11$ &   $7.1\pm 0.8$ & $5.7\pm 0.7$ & $0.67\pm 0.08$ \\
2 Southeast & $1.6\pm 0.1$ & $-121\pm 6$ & $146\pm 14$ & $0.25\pm 0.03$ & $130\pm 17$ &    $9.3\pm 1.2$ & $7.5\pm 1.0$ &  $0.88\pm 0.11$ \\
3 North & $2.9\pm 0.2$ & $162\pm 2$ & $40\pm 4$ & $0.12\pm 0.02$ & $62\pm 8$ &       $4.5\pm 0.6$ & $3.6\pm 0.5$ & $0.42\pm 0.05$ \\
\hline
\end{tabular}
\end{table}

\end{landscape}

\clearpage
\newpage

\end{document}